\documentclass[twocolumn,english,aps,preprintnumbers,amsmath,amssymb,superscriptaddress]{revtex4}
\usepackage[latin9]{inputenc}
\usepackage{amsmath}
\usepackage{graphicx}

\makeatletter
\@ifundefined{textcolor}{}
{%
 \definecolor{BLACK}{gray}{0}
 \definecolor{WHITE}{gray}{1}
 \definecolor{RED}{rgb}{1,0,0}
 \definecolor{GREEN}{rgb}{0,1,0}
 \definecolor{BLUE}{rgb}{0,0,1}
 \definecolor{CYAN}{cmyk}{1,0,0,0}
 \definecolor{MAGENTA}{cmyk}{0,1,0,0}
 \definecolor{YELLOW}{cmyk}{0,0,1,0}
}

\usepackage{dcolumn}
\usepackage{bm}
\usepackage{color}
\usepackage[final]{pdfpages}

\makeatother

\usepackage{babel}

\begin{document}

\preprint{preprint(\today)}

\title{Tunable Berry Curvature Through Magnetic Phase Competition in a Topological Kagome Magnet}







\author{Z.~Guguchia}
\email{zurab.guguchia@psi.ch} \affiliation{Laboratory for Muon Spin Spectroscopy, Paul Scherrer Institute, CH-5232 Villigen PSI, Switzerland}
\affiliation{Laboratory for Topological Quantum Matter and Spectroscopy, Department of Physics, Princeton University, Princeton, New Jersey 08544, USA}

\author{J.~Verezhak}
\affiliation{Laboratory for Muon Spin Spectroscopy, Paul Scherrer Institute, CH-5232 Villigen PSI, Switzerland}

\author{D.~Gawryluk}
\thanks{On leave from Institute of Physics, Polish Academy of Sciences, Aleja Lotnikow 32/46, PL-02-668 Warsaw, Poland}.
\affiliation{Laboratory for Multiscale Materials Experiments, Paul Scherrer Institut, 5232 Villigen PSI, Switzerland}

\author{S.S.~Tsirkin}
\affiliation{Department of Physics, University of Z\"{u}rich, Winterthurerstrasse 190, Zurich, Switzerland}

\author{J.-X.~Yin}
\affiliation{Laboratory for Topological Quantum Matter and Spectroscopy, Department of Physics, Princeton University, Princeton, New Jersey 08544, USA}

\author{I. Belopolski}
\affiliation{Laboratory for Topological Quantum Matter and Spectroscopy, Department of Physics, Princeton University, Princeton, New Jersey 08544, USA}

\author{H. Zhou}
\affiliation{International Center for Quantum Materials and School of Physics, Peking University, Beijing, China.}
\affiliation{CAS Center for Excellence in Topological Quantum Computation, University of Chinese Academy of Science, Beijing, China.}

\author{G.~Simutis}
\affiliation{Laboratory for Muon Spin Spectroscopy, Paul Scherrer Institute, CH-5232 Villigen PSI, Switzerland}

\author{S.-S. Zhang}
\affiliation{Laboratory for Topological Quantum Matter and Spectroscopy, Department of Physics, Princeton University, Princeton, New Jersey 08544, USA}

\author{T.A. Cochran}
\affiliation{Laboratory for Topological Quantum Matter and Spectroscopy, Department of Physics, Princeton University, Princeton, New Jersey 08544, USA}

\author{G. Chang}
\affiliation{Laboratory for Topological Quantum Matter and Spectroscopy, Department of Physics, Princeton University, Princeton, New Jersey 08544, USA}

\author{E.~Pomjakushina}
\affiliation{Laboratory for Multiscale Materials Experiments, Paul Scherrer Institut, 5232 Villigen PSI, Switzerland}

\author{L.~Keller}
\affiliation{Laboratory for Neutron Scattering, Paul Scherrer Institut, CH-5232 Villigen PSI, Switzerland}

\author{Z.~Skrzeczkowska} 
\affiliation{Laboratory for Multiscale Materials Experiments, Paul Scherrer Institut, 5232 Villigen PSI, Switzerland}
\affiliation{Faculty of Chemistry, Warsaw University of Technology, Noakowskiego 3, 00-664 Warsaw, Poland}  

\author{Q.~Wang}
\affiliation{Department of Physics and Beijing Key Laboratory of Opto-electronic Functional Materials and Micro-nano Devices, Renmin University of China, Beijing, China}

\author{H.C.~Lei}
\affiliation{Department of Physics and Beijing Key Laboratory of Opto-electronic Functional Materials and Micro-nano Devices, Renmin University of China, Beijing, China}

\author{R.~Khasanov}
\affiliation{Laboratory for Muon Spin Spectroscopy, Paul Scherrer Institute, CH-5232
Villigen PSI, Switzerland}

\author{A.~Amato}
\affiliation{Laboratory for Muon Spin Spectroscopy, Paul Scherrer Institute, CH-5232
Villigen PSI, Switzerland}

\author{S.~Jia}
\affiliation{International Center for Quantum Materials and School of Physics, Peking University, Beijing, China.}
\affiliation{CAS Center for Excellence in Topological Quantum Computation, University of Chinese Academy of Science, Beijing, China.}

\author{T.~Neupert}
\affiliation{Department of Physics, University of Z\"{u}rich, Winterthurerstrasse 190, Zurich, Switzerland}

\author{H.~Luetkens}
\email{hubertus.luetkens@psi.ch}
\affiliation{Laboratory for Muon Spin Spectroscopy, Paul Scherrer Institute, CH-5232 Villigen PSI, Switzerland}

\author{M.Z.~Hasan}
\email{mzhasan@princeton.edu}
\affiliation{Laboratory for Topological Quantum Matter and Spectroscopy, Department of Physics, Princeton University, Princeton, New Jersey 08544, USA}

\newpage
\maketitle

\textbf{Magnetic topological phases of quantum matter are an emerging frontier in physics and material science \cite{Keimer,WangZhang,HasanKane,Wen}. Along these lines, several kagome magnets \cite{JXYin2,LYe,THan,JXYin1,FelserCSS} have appeared as the most promising platforms. However, the magnetic nature of these materials in the presence of topological state remains an unsolved issue \cite{JXYin2,LYe,THan,JXYin1,FelserCSS}. Here, we explore magnetic correlations in the kagome magnet 
Co$_{3}$Sn$_{2}$S$_{2}$. Using muon spin-rotation, we present evidence for competing magnetic orders in the kagome lattice of this compound. Our results show that while the sample exhibits an out-of-plane ferromagnetic ground state, an in-plane antiferromagnetic state appears at temperatures above 90 K, eventually attaining a volume fraction of 
80${\%}$ around 170 K, before reaching a non-magnetic state. Strikingly, the reduction of the anomalous Hall conductivity above 90 K linearly follows the disappearance of the volume fraction of the ferromagnetic state. We further show that the competition of these magnetic phases is tunable through applying either an external magnetic field or hydrostatic pressure. Our results taken together suggest the thermal and quantum tuning of Berry curvature field via external tuning of magnetic order. Our study shows that Co$_{3}$Sn$_{2}$S$_{2}$ is a rare example where the magnetic competition drives the thermodynamic evolution of the Berry curvature field, thus tuning its topological state.}

 The kagome lattice is a two-dimensional pattern of corner-sharing triangles. With this unusual symmetry and the associated geometrical frustration, the kagome lattice can host peculiar states including flat bands \cite{JXYin1}, Dirac fermions \cite{LYe,JXYin2}  and spin liquid phases \cite{THan,SYan}. In particular, magnetic kagome materials offer a fertile ground to study emergent behaviors resulting from the interplay between unconventional magnetism and band topology. Recently, transition-metal based kagome magnets \cite{FelserCSS,Wang,THan,JXYin1,LYe,SYan,JXYin2,Nakatsuji,Nayak} are emerging as outstanding candidates for such studies,
as they feature both large Berry curvature fields and unusual magnetic tunability. In this family, the kagome magnet Co$_{3}$Sn$_{2}$S$_{2}$ is found to exhibit both a large anomalous Hall effect and anomalous Hall angle, and is identified as a promising Weyl semimetal candidate \cite{FelserCSS,Wang,QXu,Muechler}. However, despite knowing
the magnetic ground state is ferromagnetic below $T_{\rm C}$ = 177 K \cite{Kassem} with spins aligned along the $c$-axis \cite{FelserCSS,Wang,Vaqueiro} (see Figs. 1 a and b) there is no report of its magnetic tunability or phase diagram, and its interplay with the topological band structure. Here we use high-resolution ${\mu}$SR to systematically characterize the phase diagram, uncovering another intriguing in-plane antiferromagnetic phase. The magnetic competition between these two phases is further found to be highly tunable via applying either pressure \cite{GuguchiaPressure,Andreica,MaisuradzePC,GuguchiaNature} or magnetic field. Combined with first principles calculations, we discover that the tunable magnetic correlation plays a key role in determining the giant anomalous Hall transport.

\begin{figure*}[t!]
\includegraphics[width=1.0\linewidth]{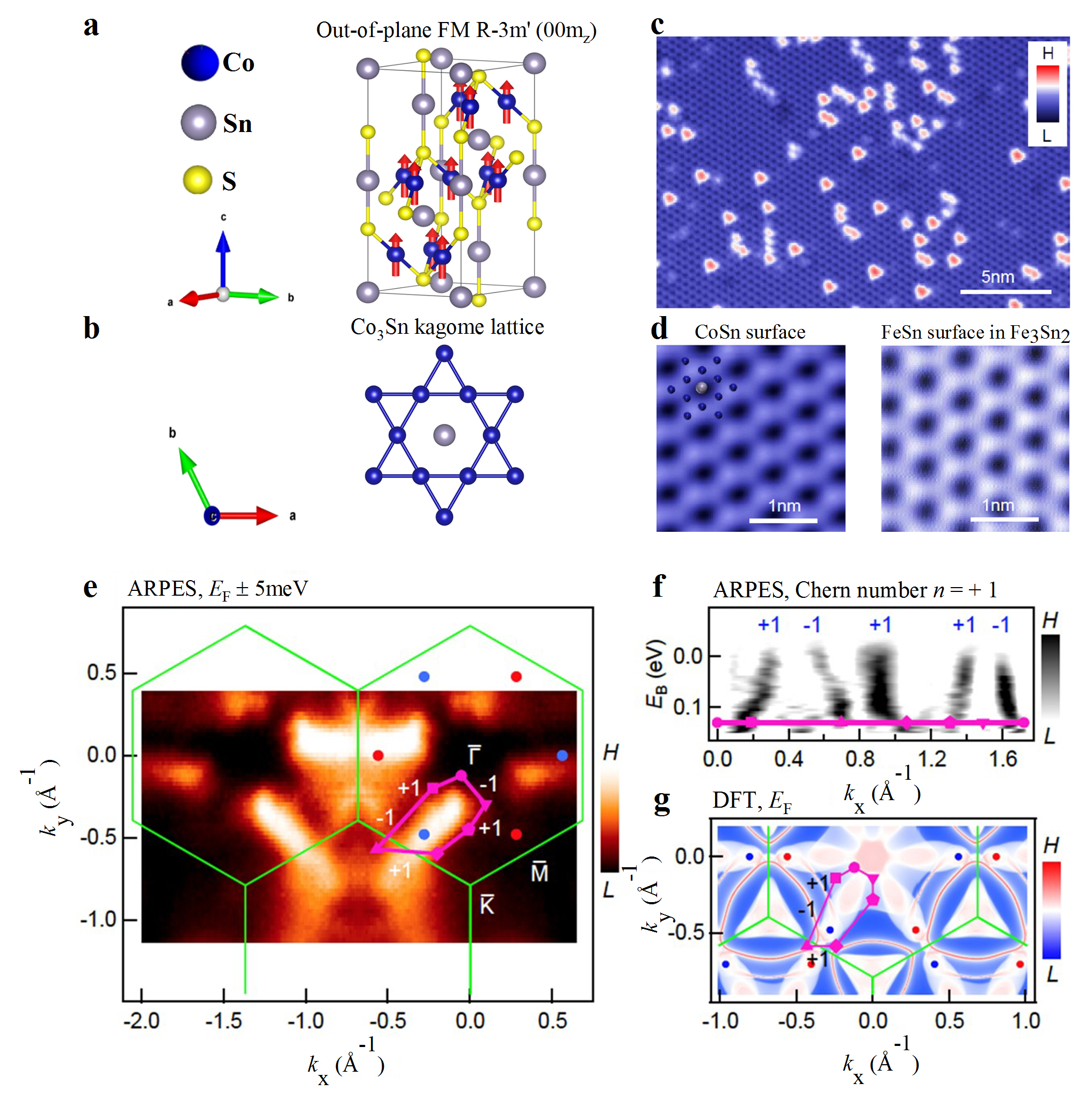}
\vspace{-1.3cm}
\caption{ (Color online) \textbf{Toplogical ground state of the kagome system Co$_{3}$Sn$_{2}$S$_{2}$.}
(a) Magnetic structure of Co$_{3}$Sn$_{2}$S$_{2}$, showing a ferromagnetic ground state with
spins on Co atoms aligned along the $c$ axis. (b) Kagome lattice structure of the Co$_{3}$Sn layer. 
(c) Topographic image of the CoSn surface. (d) A zoom-in image of the CoSn surface (left) that shows similar morphology with the FeSn surface (right) in Fe$_{3}$Sn$_{2}$. The inset illustrates the possible atomic assignment of the kagome lattice. Data are taken at the tunneling junction: $V$ = 50 mV, $I$ = 0.8 nA, $T$ = 4.2 K.
(e) Fermi surface of Co$_3$Sn$_2$S$_2$ acquired using angle-resolved photoemission spectroscopy (ARPES) at temperature $T$ = 22 K with surface Brillouin zone (green lines) determined from the crystal structure; predicted topological band crossing points (positive Chern number: blue dots, negative Chern number: red dots) from first-principles calculation; and a closed surface momentum-space loop (purple quadrilateral). (f) Energy-momentum cut along the purple loop, indicating an unconventional odd number of band crossings. 
(g) Calculated (momentum-resolved) surface density of states at $E_F$ for Co$_3$Sn$_2$S$_2$.}
\label{fig1}
\end{figure*}

Co$_{3}$Sn$_{2}$S$_{2}$ has a layered crystal structure with a CoSn kagome lattice (Fig. 1a and b).
Cleaving at cryogenic temperatures often reveals Sn and S terminated surfaces as demonstrated by our previous scanning tunnelling microscopy (STM) study \cite{JXYin1}. In addition to these two dominant surfaces, we also very rarely found CoSn surfaces (Fig 1c, left panel of d) which lies under the S surface. An enlarged view of this surface reveals a similar morphology similar to the FeSn surface in Fe$_{3}$Sn$_{2}$ at the atomic level \cite{JXYin2}, both of which are consistent with the transition metal based kagome lattice structure as seen in the STM images in Fig 1d. Having confirmed the fundamental kagome lattice in the crystal structure, we further characterize its electronic structure by angle-resolved photoemission spectroscopy (ARPES). We study the Fermi surface at a temperature of 22 K and observe regions of broad spectral weight suggesting bulk states, interspersed with sharper features suggesting surface states (Fig. 1e). Based on fundamental theoretical considerations, the electronic structure of a ferromagnet generically hosts Weyl fermions in the bulk, associated with a non-zero Chern number. Furthermore, for a closed momentum-space path in the surface Brillouin zone, the net number of surface states on the path equals the net enclosed bulk Chern number \cite{HasanWeyl,Vishwanath-Review,IlyaHasan}. Motivated by these considerations, we study the ARPES dispersion on a closed loop enclosing the observed bulk pocket (purple contour in Fig. 1e). We observe five crossings at the Fermi level, inconsistent with any conventional surface state band structure (Fig. 1f). We next sum up the net number of crossings on the loop, associating a $+1$ with left-propagating modes and a $-1$ with right-propagating modes. We find a net value of $+1$, suggesting that the loop encloses a Chern number of $n = +1$ and implying that at least one Weyl fermion lies inside the loop. To more deeply understand our result, we calculate the distribution of Weyl points from DFT and we observe that one Weyl point projects inside our closed loop (blue dot in Fig. 1e), consistent with our observation of a Chern number of $n = +1$ using ARPES. We next calculate the entire Fermi surface and we observe that the Weyl points are connected pairwise by Fermi arc surface states, along with trivial surface state pockets enclosing the $\bar{K}$ and $\bar{K}'$ points (Fig. 1g, see also Supplementary Information). We can understand the crossings observed in ARPES by associating them with these surface states in DFT. Proceeding from left to right in Fig. 1f, we see that the first crossing ($+1$) is the Fermi arc, the next two ($-1$ and $+1$) are associated with the $\bar{K}$ point surface state, while the last two ($+1$ and $-1$) may be associated with a surface resonance arising from bulk pockets projecting near $\bar{\Gamma}$. In this way, the correspondence between ARPES and DFT allows us to identify the first chiral mode in Fig. 1f as the topological Fermi arc surface state. Our experimental observations suggest the existence of a topologically non-trivial band structure in the kagome magnet Co$_3$Sn$_2$S$_2$.

\begin{figure*}[t!]
\includegraphics[width=1.0\linewidth]{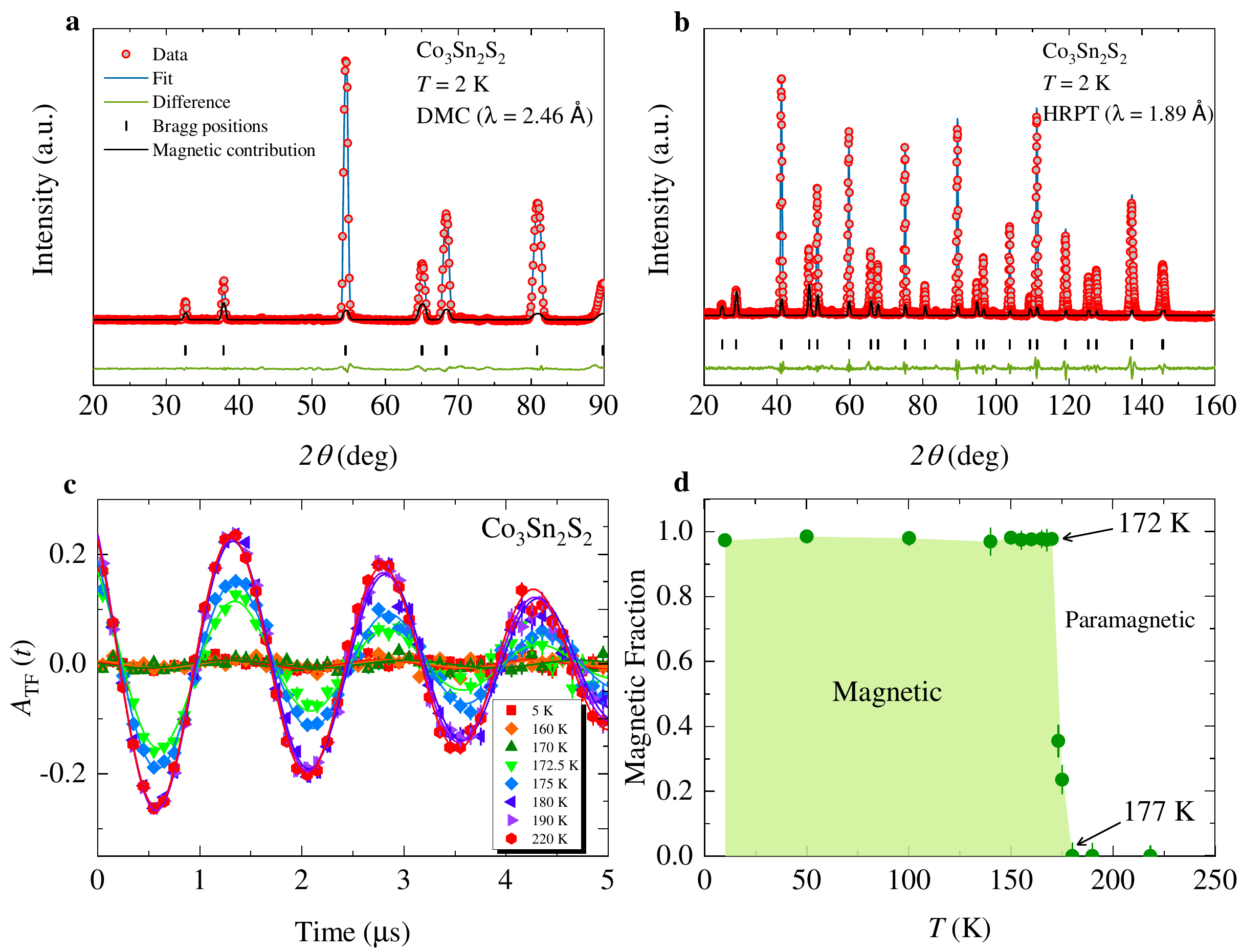}
\vspace{-0.5cm}
\caption{ (Color online) \textbf{Ferromagnetic ground state and the temperature evolution of the magnetically ordered volume fraction in the kagome system Co$_{3}$Sn$_{2}$S$_{2}$.}
(a-b) Neutron powder diffraction pattern, recorded at 2 K for the sample Co$_{3}$Sn$_{2}$S$_{2}$ with two different instruments. The solid black lines represent a Rietveld refinement profile. The residuals are plotted at the bottom of the figure. The solid green lines are the fitted magnetic contributions. To better visualise the magnetic peaks, the intensities are multipled by 500 and 300 for (a) and (b), respectively. (c) The weak-TF ${\mu}$SR spectra, obtained above and below Curie temperature $T_{\rm C}$. (d) The temperature dependence of the magnetically ordered volume fraction, extracted from the amplitude of the TF ${\mu}$SR spectra.}
\label{fig1}
\end{figure*}

 Having characterized the topologically non-trivial band structure, we probe the crystal and magnetic structure of Co$_{3}$Sn$_{2}$S$_{2}$ at base temperature  (Figs. 2a and b).  The crystal structure of Co$_{3}$Sn$_{2}$S$_{2}$ through the entire temperature range was well refined with Rietveld refinements of the raw neutron diffraction data, employing a rhombohedral lattice structure in the space group R-3m. An example of the refinement profile for the 2 K data is shown in Fig. 2a and b, and no secondary phase can be detected. We do not observe any substantial additional diffraction intensity at 2 K in comparison with the paramagnetic state. Instead, it appears that the only clearly visible effect on the Bragg peak intensities originates from the decrease in the Debye-Waller atomic displacement parameters. To estimate the upper limit of the magnetic moment at the Co sites as well as to identify the magnetic structure, we have performed a group theoretical analysis, described in detail below, and a Rietveld refinement of the data. Since we know from the magnetic susceptibility data that the sample is out-of-plane ferromagnetically ordered \cite{Kassem}, the minimal realistic model to refine at 2 K is with the R-3m$'$ structure. 
By including only the ferromagnetic (FM) $z$-component in the refinement we deduce $m_{z}$ = 0.269(102) ${\mu_B}$. By including all parameters in the refinement, as well as the atomic displacements (ADP) we obtain a magnetic moment, $\bf{M}$ = ($m_{x}$,$m_{y}$,$m_{z}$) = [0.093(80), 0.187(160), 0.255(137)] ${{\mu}_B}$. Thus the upper estimate of the long range ordered magnetic moment per Co at base temperature is 0.2 -- 0.3 ${\mu_B}$. Our neutron diffraction data support the $z$-axis FM order in Co$_{3}$Sn$_{2}$S$_{2}$ and provides an estimate for the upper bound of the ordered moment.

 Next we study the magnetism and its temperature, pressure and field dependence in Co$_{3}$Sn$_{2}$S$_{2}$ using the ${\mu}$SR technique, which serves as an extremely sensitive local probe for detecting small internal magnetic fields and ordered magnetic volume fractions in the bulk of magnetic materials. The weak transverse field (weak-TF) ${\mu}$SR spectra and the temperature dependence of the magnetically ordered volume fraction for Co$_{3}$Sn$_{2}$S$_{2}$ are shown in Fig. 2c and d, respectively. In a weak-TF, the amplitude of the low-frequency oscillations is proportional to the paramagnetic volume fraction. Thus, a spectrum with no oscillation corresponds to a fully ordered sample, while a spectrum with oscillation in the full asymmetry indicates that the sample is in a paramagnetic state. The weak-TF spectra below $T$ ${\sim}$ 172 K show negligibly small 3 ${\%}$ oscillation amplitudes, demonstrating a fully magnetically ordered ground state. Above 172 K, the oscillation amplitude increases and reaches the full paramagnetic volume at $T_{\rm C}$ ${\simeq}$ 177 K. The temperature dependent magnetic fraction therefore shows a relatively sharp transition from the paramagnetic to the magnetic state with the coexistence of magnetic and paramagnetic regions in the temperature interval 
172 K -- 177 K, i.e., only very close to the transition. 

\begin{figure*}[t!]
\includegraphics[width=1.0\linewidth]{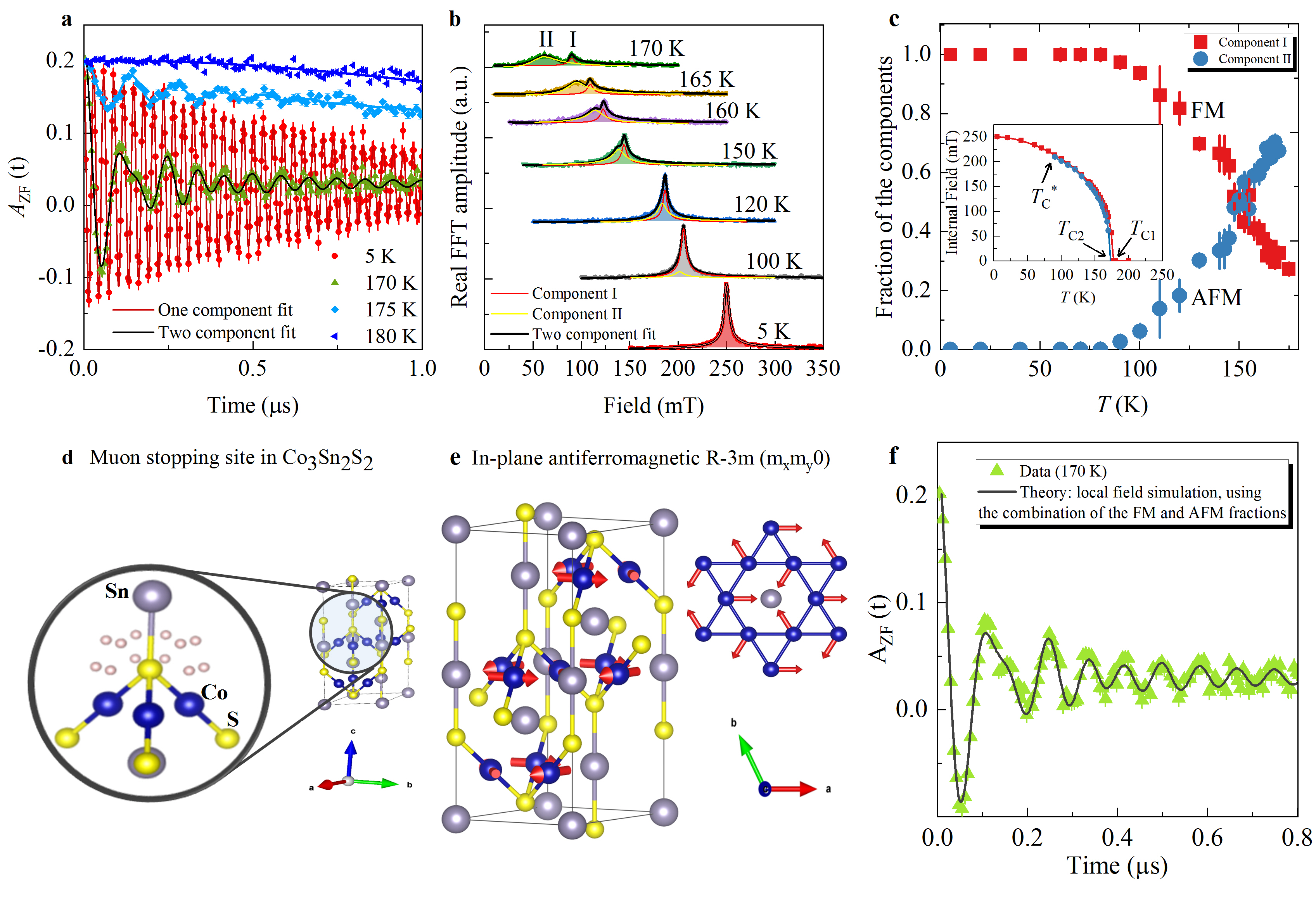}
\vspace{-1.0cm}
\caption{ (Color online) \textbf{Macroscopic phase separation between two distinct magnetically ordered regions in Co$_{3}$Sn$_{2}$S$_{2}$.}
(a) Zero field spectra, recorded at temperatures above and below $T_{\rm C}$. The solid lines are the fit of the data using Eq. 2. (b) Fourier transform amplitudes of the oscillating components of the ${\mu}$SR time spectra as a function of temperature. (c) The temperature dependences of the relative volume fractions of the two magnetically ordered regions. Inset shows the temperature dependences of the internal magnetic fields for the two components. Arrows mark the critical temperatures $T_{\rm C1}$ and $T_{\rm C2}$ for high frequency and low frequency components, respectively as well as the transition temperature $T_{\rm C}^{*}$, below which only one component signal is observed. (d) Crystallographically equivalent muon stopping sites with the Wyckoff position 36i within the structure of CoSn$_{2}$S$_{2}$. (e) In-plane antiferromagnetic structure. Gray solid lines indicate the boundaries of a single unit cell of the crystal structure. (f) ${\mu}$SR spectrum, recorded at 170 K, is shown along with the result of the local field simulation (black solid line) at the muon stopping site, considering two distinct magnetically ordered regions with out-of-plane and in-plane magnetic configurations, respectively.}
\label{fig1}
\end{figure*}

 In order to study the detailed temperature evolution of the magnetic order parameter in Co$_{3}$Sn$_{2}$S$_{2}$, zero-field ${\mu}$SR measurements were carried out. Time-spectra, recorded below (5 K, 170 K, and 175 K) and above (180 K) the magnetic ordering temperature, are shown in Fig. 3a. At $T$ = 180 K, the entire sample is in the paramagnetic state as evidenced by the weak ${\mu}$SR depolarization and its Gaussian functional form arising from the interaction between the muon spin and randomly oriented nuclear magnetic moments \cite{Toyabe}. At $T$ = 5 K, a spontaneous muon spin precession with a well-defined single frequency is observed, which is clearly visible in the raw data (Fig. 3a). In addition, the spectrum is characterised by the low value of the transverse muon spin depolarization rate ${\lambda}_T$ = 1.2(1) ${\mu}s^{-1}$, which is a measure of the width of the static magnetic field distribution at the muon site, implying a narrow field distribution in the sample and thus a very homogeneous magnetic ground state. Our stopping site calculations and stability analysis reveal one plausible muon stopping site to be the Wyckoff position 36i (see Fig. 3d) in Co$_{3}$Sn$_{2}$S$_{2}$, which is consistent with the observation of only one precession frequency in the ZF-${\mu}$SR spectra at base temperature [details on the muon stopping site and local field calculations can be found in the supplementary information]. The single frequency behaviour is robust up to ${\sim}$ 90 K, after which a second frequency begins to appear (Fig. 3a).

To better visualize this effect, we show the Fourier transform amplitudes of the oscillating components of the ${\mu}$SR time spectra as a function of temperature (Fig. 3b). Starting from $T$ = 90 K, two distinct precession frequencies 
appear in the ${\mu}$SR spectra, which can be clearly seen and are well separated when approaching the transition temperature. The temperature dependences of the internal fields (${\mu}_{0}$$H_{int}$ = ${\omega}$/${\gamma}_{\mu}$$^{-1}$) for the two components are shown in the inset of Fig. 3c. Both order parameters show the monotonous decrease and clear separation  with increasing temperature. It is important to note that the two components have  slightly different transition temperatures, with the high frequency component having an onset at  $T_{\rm {C1}}$ ${\simeq}$ 177 K, and the low frequency component having an onset of $T_{\rm {C2}}$ ${\simeq}$ 172 K. Remarkably, the low frequency component develops at the cost of the high frequency component, since the appearance of the low frequency component above  $T_{\rm C}^{*}$ ${\sim}$ 90 K is accompanied by the reduction of the volume fraction of the high frequency one. As shown in Fig. 3c, the fraction of the high frequency component continuously decreases with increasing the temperature above  $T_{\rm C}^{*}$ ${\sim}$ 90 K and close to the transition the low frequency component acquires the majority of the volume. These results suggest the presence of two distinct magnetically ordered regions in Co$_{3}$Sn$_{2}$S$_{2}$ at temperatures above ${\sim}$ 90 K. Note that the disappearance of the low frequency component above $T_{\rm {C2}}$ ${\simeq}$ 172 K explains the reduction of the total magnetic fraction in the temperature interval between $T_{\rm {C2}}$ ${\simeq}$ 172 K and $T_{\rm {C1}}$ ${\simeq}$ 177 K (see Fig. 2d). 

\begin{figure*}[t!]
\includegraphics[width=1.0\linewidth]{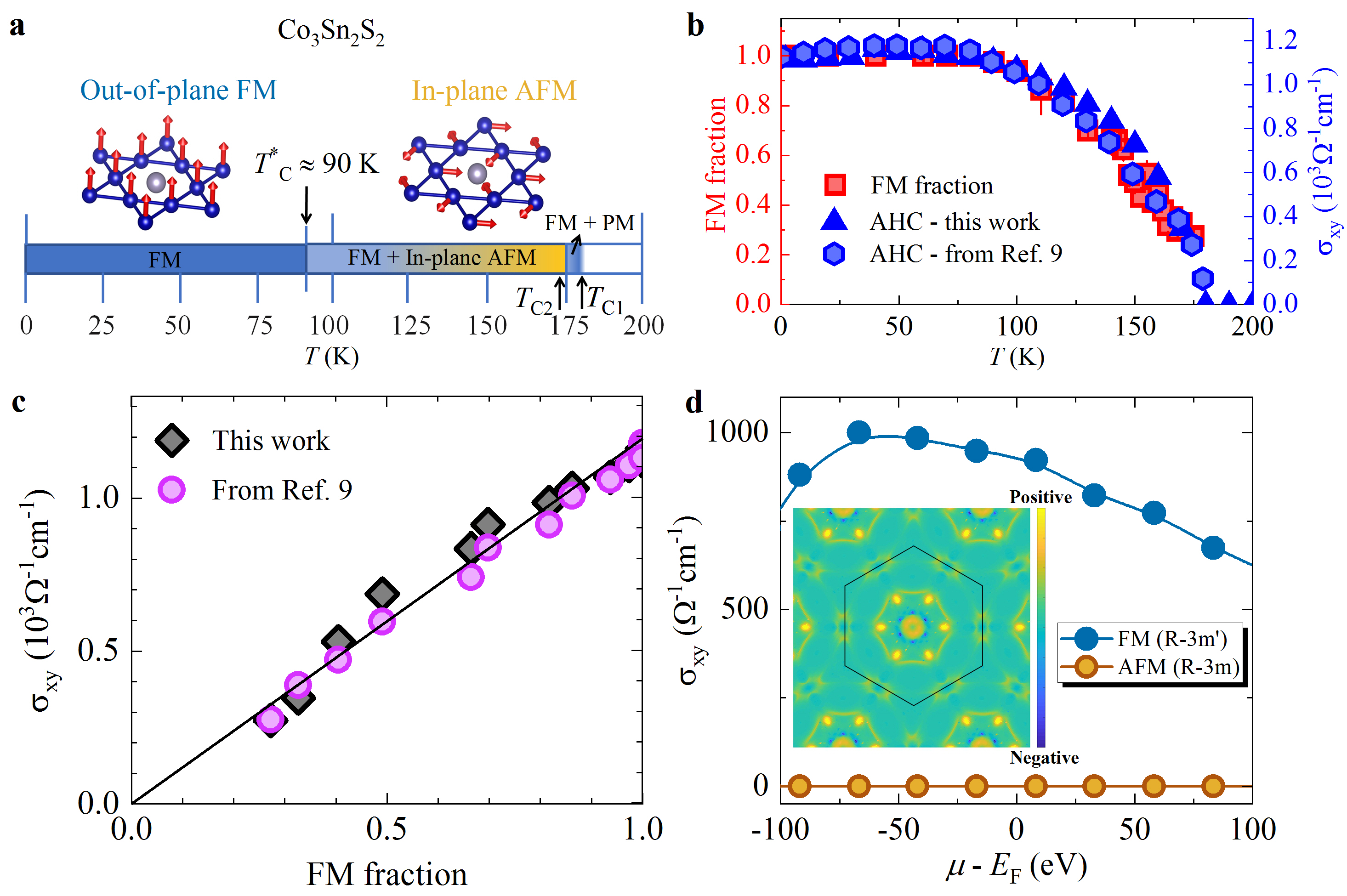}
\vspace{-0.5cm}
\caption{ (Color online) \textbf{Magnetic competition driven thermal evolution of anomalous Hall conductivity.}
(a) Schematic magnetic phase diagram as a function of temperature and spin structures of Co$_{3}$Sn$_{2}$S$_{2}$, i.e. the FM and the in-plane AFM structures. The arrows mark the transition temperatures $T_{\rm {C1}}$ 
${\simeq}$ 177 K, $T_{\rm {C2}}$ ${\simeq}$ 172 K and $T_{\rm C}^{*}$ ${\simeq}$ 90 K. (b) The temperature dependence of the fraction of the FM volume fraction and the in-plane anomalous Hall conductivity ${\sigma}_{xy}$. (c) The correlation plot of ${\sigma}_{xy}$ vs fraction of the ferromagnetically ordered region. The solid straight line is drawn between a hypothetical situation of the minimum (zero) and the maximum values of ${\sigma}_{xy}$ and the FM fraction. Data shown in solid circles are taken from Ref. \cite{FelserCSS}. (d) The dependence of ${\sigma}_{xy}$ on the chemical potential, calculated for the out of plane FM and the in-plane AFM structures. The inset shows the calculated Berry curvature distribution in the BZ at the Ferromagnetic phase.}
\label{fig1}
\end{figure*}

\begin{figure*}[t!]
\includegraphics[width=1.0\linewidth]{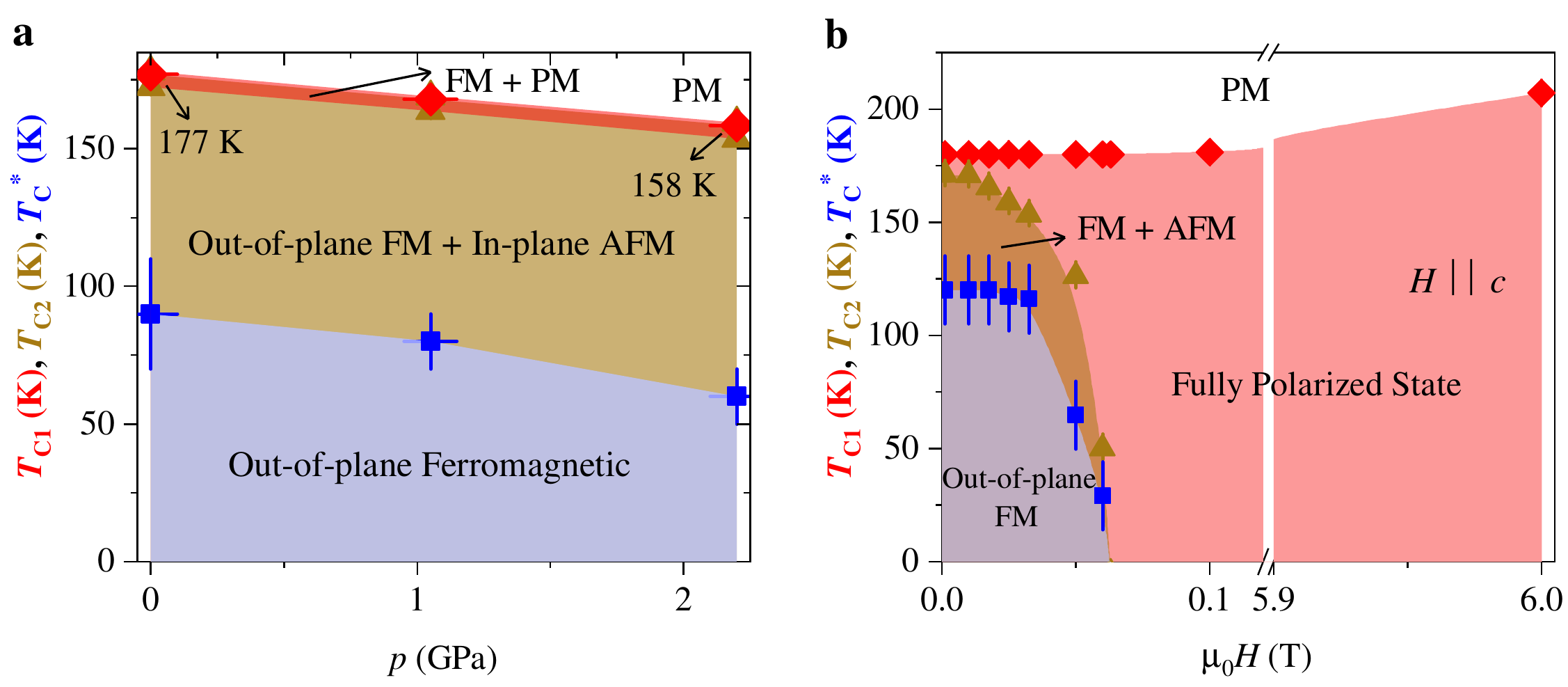}
\vspace{-0.5cm}
\caption{ (Color online) \textbf{Magnetic field and hydrostatic pressure tuning of the magnetic competition in Co$_{3}$Sn$_{2}$S$_{2}$.}
(a) Temperature-Pressure phase diagram for Co$_{3}$Sn$_{2}$S$_{2}$ obtained by microscopic ${\mu}$SR technique. (b) Temperature-Field phase diagram for Co$_{3}$Sn$_{2}$S$_{2}$, obtained from the macroscopic magnetisation measurements.}
\label{fig1}
\end{figure*}

   To understand the multi-domain physics further, we investigate the possible magnetic structures in Co$_{3}$Sn$_{2}$S$_{2}$. According to previously reported magnetisation measurements, Co$_{3}$Sn$_{2}$S exhibits out-of-plane FM ordering below $T_{\rm {C1}}$ ${\simeq}$ 177 K. The FM structure forces the propagation vector to be $\bf{k}$ = (0,0,0), so we consider the maximally symmetric Shubnikov magnetic subgroups of the grey paramagnetic group R-3m1$'$ for the ${\Gamma}$-point of the Brillouin zone. The decomposition of the magnetic representation for Co in Wyckoff position 9d (1/2,0,1/2) reads ${\Gamma}$$_{1}$$^{+}$ ${\otimes}$ 2${\Gamma}$$_{2}$$^{+}$ ${\otimes}$ 3${\Gamma}$$_{3}$$^{+}$ (${\tau}_{\rm 7}$) with the dimensions of irreproducible representations (irreps) one, one, and two, respectively. The first two irreps result in the R-3m$'$ and R-3m subgroups, which are the solutions of maximal symmetry with the smallest number of the refined parameters. The first subgroup R-3m$'$ has two spin components $\bf{M}_{\rm Co}$ = ($m_{\rm x}$,2$m_{\rm x}$,$m_{\rm z}$) with AFM in the plane and FM along the $c$-axis. Previous reports \cite{Vaqueiro} as well as our neutron experiments suggest a $c$-axis aligned  R-3m$'$ FM structure (see Fig. 1a) (although some canting away from the $c$-axis is possible) as the magnetic ground state for Co$_{3}$Sn$_{2}$S$_{2}$. The second subgroup R-3m is 120$^{\circ}$ antiferromagnetic (AFM) order with the spins in ($ab$) plane (see Fig. 3e), which we will refer to as in-plane AFM order. In addition to the above symmetry considerations, we also investigate the magnetic structure using density functional theory (DFT). We find that among the two magnetic arrangements as shown in Fig. 1a and 3e, the lowest energy configuration is FM with the magnetic moments along the $c$-axis, with magnitude ${\sim}$ 0.35 ${\mu_B}$/atom. The in-plane AFM configuration with the same magnetic moment has a higher energy of ${\sim}$ 20 meV/Co atom, relative to the $c$-axis FM order. Given the information of the allowed magnetic structures and the muon stopping site in Co$_{3}$Sn$_{2}$S$_{2}$, the local fields were calculated for both out-of-plane FM and in-plane AFM spin arrangements at the single muon site with the Wyckoff position 36i (see Fig. 3d) and then fit with the fractions and moment sizes for each phase. The result of the calculation for 170K is shown in Fig. 3f as the black dashed line, which lines up with the experimental data very well. Moreover, the temperature dependence of the moment sizes, relative fractions of two magnetic components and the relaxation rates (see supplementary information) also agree very well with the experimental data. Thus, the analysis confirms the two magnetically ordered fractions in Co$_{3}$Sn$_{2}$S$_{2}$ with different moment sizes: at low temperatures the out-of-plane FM structure is dominant and  with increasing temperature the fraction of the in-plane AFM state grows, and becomes the dominant component at 170 K.


 The schematic magnetic phase diagram and spin structures of Co$_{3}$Sn$_{2}$S$_{2}$ are drawn in Fig. 4a, showing the evolution of the out-of-plane FM and in-plane AFM states as a function of temperature. Our key finding is the observation of a phase separated magnetically ordered region in the temperature interval of 
$T_{\rm C}^{*}$ ${\textless}$ $T$ ${\textless}$ $T_{\rm {C2}}$ in the magnetic Weyl semimetal candidate Co$_{3}$Sn$_{2}$S$_{2}$.  This conclusion is robust and model-independent, as it relies on the extreme sensitivity of ${\mu}$SR, which is a local probe, to ordered magnetic volume fractions in the bulk of magnetic materials. Neutron diffraction measurements provide an estimate of the upper bound of the long range ordered magnetic moment 0.2 -- 0.3 ${\mu_B}$ per Co at base temperature.  Local field simulations at the muon stopping site further shed light on the possible magnetic states. For $T$ ${\textless}$ $T_{\rm C}^{*}$ = 90 K, the entire sample exhibits an out-of-plane FM structure. When $T$ ${\geq}$ 90 K, the in-plane AFM arrangement appears, and its volume fraction grows with increasing temperature and eventually dominates around 170 K, before it disappears at $T_{\rm {C2}}$ = 172 K. Above $T_{\rm {C2}}$, the sample exhibits the small volume fraction with the out-of-plane FM order (the rest of the volume is occupied by the paramagnetic state) until $T_{\rm {C1}}$ = 177 K, before reaching a fully paramagnetically ordered state. According to our DFT calculations the energies of the out-of-plane FM and in-plane AFM configurations are similar. Thermodynamics may affect this intricate balance, and tip it in favour of the in-plane magnetic structure. The interplay between different magnetically ordered regions, each of which may possess distinct topological invariants, can possibly give rise to exciting physics at the magnetic domain boundaries. 

The ${\mu}$SR observation of the presence of the novel magnetic phase for $T_{\rm C}^{*}$ ${\textless}$ $T$ ${\textless}$ $T_{\rm {C2}}$ in Co$_{3}$Sn$_{2}$S$_{2}$ is also supported by the temperature dependent magnetic susceptibility data (see supplementary information), which shows the broad maximum within that temperature range. However, we note that the anomaly in susceptibility appears at ${\sim}$ 125 K, which is higher than $T_{\rm C}^{*}$ ${\sim}$ 90 K observed by ${\mu}$SR. This is most likely related to the fact that the volume fraction of the new magnetic phase is small at 90 K and gradually increases with increasing temperature. Since the microscopic ${\mu}$SR technique is extremely sensitive to small volume fractions, it allows it to probe the true onset temperature of the transition. Macroscopic magnetization measurements detect the transition at higher temperatures, where the magnetic phase acquires a high enough volume fraction.

 One of the most striking effects in Co$_{3}$Sn$_{2}$S$_{2}$ is a large intrinsic anomalous Hall conductivity (AHC) and a giant anomalous Hall angle, due to the considerably enhanced Berry curvature arising from its band structure \cite{FelserCSS}. 
In order to explore the correlation between the novel magnetic properties found in this work and the topological aspects of Co$_{3}$Sn$_{2}$S$_{2}$, we compare the temperature dependence of the anomalous Hall conductivity to the temperature evolution of the volume fraction of the out-of-plane FM component (Fig. 4b). Remarkably, the temperature evolution of the AHC matches very well with the evolution of the FM ordered volume fraction (Fig. 4b and c). The magnitude of the AHC is robust against the increase in temperature up until $T$ ${\sim}$ 90 K, after which it gradually decreases, corresponding to the temperature at which the in-plane AFM volume fraction begins to increase. This is, to our knowledge, the first example of such an excellent correlation between the magnetic volume fraction and the Berry curvature induced AHC. From symmetry arguments, we can extract important clues of which components of the conductivity tensor ${\sigma}$ are forbidden and which are allowed to be nonzero. Due to the 3-fold rotational symmetry the ${\sigma}_{xz}$ and ${\sigma}_{yz}$ components are zero in all considered magnetic structures. The in-plane AFM R-3m structure, shown in Fig. 3e, has a 2-fold rotational symmetry with an axis in the $ab$ plane. Such a rotation transforms ${\sigma}_{xy}$ to -${\sigma}_{xy}$, and hence forces it to be zero (Fig. 4d). On the other hand, for the $c$-axis R-3m$'$ FM ordering, shown in Fig. 1a, we obtain an extremely high ${\sigma}_{xy}$ = 10$^{3}$ ${\Omega}$$^{-1}$cm$^{-1}$ (Fig. 4d). We can thus conclude from first principles calculations that the AHC is dominated by the FM $c$-axis component, providing an explanation for the reduction of the AHC when the ordered volume fraction of the out-of-plane FM state decreases. These observations demonstrate for the first time volume-wise magnetic competition in Co$_{3}$Sn$_{2}$S$_{2}$ as well as a remarkable correlation between the magnetically ordered fraction and the AHC.

 For further insight into the magnetic order and the novel magnetic phase transition in Co$_{3}$Sn$_{2}$S$_{2}$, ZF ${\mu}$SR experiments were carried out as a function of hydrostatic pressure. From the pressure dependent data (see the supplementary information), we can construct a temperature-pressure phase diagram for Co$_{3}$Sn$_{2}$S$_{2}$. The pressure dependences of the transition temperatures $T_{\rm {C1}}$, $T_{\rm {C2}}$, and $T_{\rm C}^{*}$ are shown in Fig. 5a. We find that hydrostatic pressure has a significant effect on the magnetic properties of this material. Namely, increased pressure results in a substantial reduction of $T_{\rm {C1}}$, $T_{\rm {C2}}$  and $T_{\rm C}^{*}$ while the fully magnetically ordered volume fraction remains intact. We also find that  under pressure, the low frequency component arising from the AFM phase appears at a lower temperature in comparison to ambient pressure. This implies that increased pressure tends to stabilise the high temperature in-plane AFM structure. These findings show that one can physically tune the magnetism in these materials with pressure. We also constructed a temperature-field phase diagram for Co$_{3}$Sn$_{2}$S$_{2}$ based on the bulk magnetization data, where the anomalies related to the ferromagnetic and the novel magnetic transitions are clearly observed (see  supplementary information). Figure 5b shows the field dependences of the critical temperatures $T_{\rm {C1}}$, $T_{\rm {C2}}$ and $T_{\rm C}^{*}$. The phase diagram implies that relatively low fields are enough to tune the magnetic state in Co$_{3}$Sn$_{2}$S$_{2}$.

The above results show that one can tune the competition between FM and AFM states and have control of the AHC by varying the temperature or varying a non-thermal parameter such as pressure and field. We note that the change in the AHC represents a change in the Berry curvature of the occupied states. The change comes from the modifications of the band dispersion or of the Berry curvature of the occupied states themselves. Thus, these findings provide us with the unique possibility of a thermal or quantum tuning of Berry curvature in Co$_{3}$Sn$_{2}$S$_{2}$ through tuning the magnetic competition.

The exploration of topological electronic phases that result from strong electronic correlations is a frontier in condensed matter physics. Kagome lattice systems are an ideal setting in which strongly correlated topological electronic states may emerge. The simplest Bloch band structure on the kagome lattice naturally includes a flat band \cite{JXYin1}, in which the effects of electronic correlations are enhanced. Our key finding is to establish Co$_{3}$Sn$_{2}$S$_{2}$ as a material that hosts topological electronic states and frustrated magnetism. Our experiments suggest that the Co spins have both ferromagnetic interactions along $c$-axis and antiferromagnetic interactions within the kagome plane, and there is a temperature dependent competition between these two ordering tendencies. The interplay between this intricate magnetism and the spin-orbit coupled band structure further induces non-trivial variations of its topological properties, which is characterized by a striking correlation between the anomalous Hall conductivity and the ferromagnetic volume fraction. Our results demonstrate thermal and quantum tuning of Berry curvature mediated by changes in the frustrated magnetic structure.

\section{METHODS}

\textbf{General remarks}: We concentrate on the high resolution, high field and high pressure \cite{GuguchiaPressure,Andreica,MaisuradzePC,GuguchiaNature} muon spin relaxation/rotation (${\mu}$SR) measurements of the temperature, pressure and field dependence of the magnetic moment as well as the magnetically ordered volume fraction in 
Co$_{3}$Sn$_{2}$S$_{2}$ and the high resolution neutron powder diffraction measurements of the tentative magnetic structure at the base temperature.  Muon stopping site and local field calculations are carried out to gain insights into the ${\mu}$SR results. In a ${\mu}$SR experiment, positive muons implanted into a sample serve as an extremely sensitive local probe to detect small internal magnetic fields and ordered magnetic volume fractions in the bulk of magnetic materials. Angle-resolved photoemission spectroscopy (ARPES) measurements and density functional theory calculations were used to explore the electronic band structure.
Neutron diffraction has the ability to directly measure the magnetic structure. The techniques of ${\mu}$SR, ARPES and DFT complement each other ideally as we are able to study the detailed temperature dependence of the magnetic order parameter and ordered volume fractions with ${\mu}$SR experiments, and correlate them with the electronic structure measured and calculated by ARPES and DFT, respectively.\\

\textbf{Sample preparation}: Stoichiometric single crystals have been grown by a modified vertical Bridgman technique as described elsewhere \cite{Holder}. The corresponding polycrystalline Co$_{3}$Sn$_{2}$S$_{2}$ were prepared by the solid state reaction method. A stoichiometric ratio of Co (2N+), Sn (2N8) and S (5N) was weighted carefully mixed, and pressed into a pellet. The sample was then evacuated and sealed in double wall quartz ampoule. The sample was slowly heated, with a rate 15$^{\rm o}$C/h, up to 450$^{\rm o}$C and pre-annealed for 24 h. Subsequently, the sample was heated up to 700$^{\rm o}$C, kept at this temperature for 72 h and cooled down to room temperature with a rate 400$^{\rm o}$C/h. The whole procedure, with exception of pre-heating, was repeated three times with intermediate re-grounding and pelletizing in a glovebox. To avoid any contamination from the environment the synthesized materials were handled in the helium filled glovebox.\\

\textbf{Pressure cell}:  Pressures up to 2.2 GPa were generated in a double wall piston-cylinder
type of cell made of CuBe/MP35N material, especially designed to perform ${\mu}$SR experiments under
pressure \cite{GuguchiaPressure,Andreica,MaisuradzePC,GuguchiaNature}. A pressure transmitting medium Daphne oil was used. The pressure was measured by tracking the SC transition of a very small indium plate by AC susceptibility. The filling factor of the pressure cell was maximized. The fraction of the muons stopping in the sample was approximately 40 ${\%}$.\\

\textbf{Neutron powder diffraction experiments}: The magnetic and crystal structures in Co$_{3}$Sn$_{2}$S$_{2}$ have been studied by neutron powder diffraction (NPD) in the temperature range 2 -- 250K. The NPD experiments were carried out at the Swiss Spallation Neutron Source (SINQ), Paul Scherrer Institute, Villigen, Switzerland. Approximately 2.5 g of powder sample was loaded into a 6 mm diameter vanadium can. The diffraction patterns were recorded between 1 and 250 K at the cold neutron powder diffractometer (DMC) \cite{Schefer} using ${\lambda}$ = 2.4576  ${\AA}$  (pyrolytic graphite (002), 2${\theta}_{max}$ = 92.7 $^{\rm o}$, 2${\theta}_{step}$ = 0.1$^{\rm o}$) and at the High-Resolution Powder diffractometer for Thermal neutrons (HRPT) \cite{Fischer} using ${\lambda}$ = 1.154, 1.8857, and 2.449  {\AA} [Ge (822), 2${\theta}_{max}$ = 160$^{\rm o}$, 2${\theta}_{step}$ = 0.05$^{\rm o}$]. Large statistics acquisitions for magnetic structure refinements were made at both 2 K and 170 K. The refinements of the crystal structure parameters were done using FullProf suite \cite{Rodriguez}, with the use of its internal tables for neutron scattering lengths. The symmetry analysis was performed using ISODISTORT tool based on ISOTROPY software, \cite{Stokes,Campbell} BasiRep program \cite{Rodriguez} and software tools of the Bilbao crystallographic server \cite{Aroyo}.\\

\textbf{${\mu}$SR experiment}: In a ${\mu}$SR experiment nearly 100 ${\%}$ spin-polarized muons (${\mu}$$^{+}$)
are implanted into the sample one at a time. The positively
charged ${\mu}$$^{+}$ thermalize at interstitial lattice sites, where they
act as magnetic microprobes. In a magnetic material the
muon spin precesses in the local field $B_{\rm \mu}$ at the
muon site with the Larmor frequency ${\nu}_{\rm \mu}$ = $\gamma_{\rm \mu}$/(2${\pi})$$B_{\rm \mu}$ [muon
gyromagnetic ratio $\gamma_{\rm \mu}$/(2${\pi}$) = 135.5 MHz T$^{-1}$].

The low background GPS (${\pi}$M3 beamline) instrument was used to study the single crystalline as well as the polycrystalline samples of Co$_{3}$Sn$_{2}$S$_{2}$ at ambient pressure.
${\mu}$SR experiments under pressure were performed at the ${\mu}$E1 beamline of the Paul Scherrer Institute (Villigen, Switzerland), where an intense high-energy ($p_{\mu}$ = 100 MeV/c) beam of muons is implanted in the
sample through the pressure cell. Transverse-field (TF) ${\mu}$SR experiments were performed at the ${\pi}$E3 beamline
of the Paul Scherrer Institute (Villigen, Switzerland), using the HAL-9500 ${\mu}$SR spectrometer. The specimen was
mounted in a He gas-flow cryostat with the $c$-axis parallel to the muon beam direction, along which the external field was applied. Magnetic fields between
10 mT and 8 T were applied, and the temperatures were varied between 3 K and 300 K.\\

\textbf{Analysis of Weak TF-${\mu}$SR data}: The wTF asymmetry spectra were analyzed by the function \cite{AndreasSuter}:

\begin{equation}
\begin{aligned}A_S(t)= A_{p}\exp(-\lambda t)\cos(\omega t + \phi),
\label{eq1}
\end{aligned}
\end{equation}
where $t$ is time after muon implantation, $A$($t$) is the time-dependent asymmetry, $A_{p}$
is the amplitude of the oscillating component (related to the paramagnetic volume
fraction), ${\lambda}$ is an exponential damping rate due to paramagnetic spin fluctuations
and/or nuclear dipolar moments, ${\omega}$ is the Larmor precession frequency set, which is proportional to the
strength of the transverse magnetic field, and ${\phi}$ is a phase offset. As it is standard for the analysis of wTF data from magnetic samples the zero for $A$($t$) was allowed to vary for each temperature to deal with the asymmetry baseline shift known to occur for magnetically ordered samples. From these
refinements, the magnetically ordered volume fraction at each temperature $T$ was determined by 
1  -- $A_{p}(T)$/$A_{p}(T_{max})$, where $A_{p}$($T_{max}$) is the amplitude in the paramagnetic phase at high temperature.\\

\textbf{Analysis of ZF-${\mu}$SR data}: All of the ZF-${\mu}$SR spectra were fitted using the following model \cite{AndreasSuter}:
	
	
	\begin{widetext}
		\begin{equation}
		\label{eq:ZFPolarizationFit}
		A_{\textrm{ZF}}(t) = F \left( \sum_{j = 1}^{2}  \left( f_j \cos(2\pi\nu_j t + \phi) e^{-\lambda_j t} \right) +
		f_L e^{-\lambda_{L} t} \right) + (1-F) \left(\frac{1}{3} + \frac{2}{3}\left( 1 - (\sigma t)^2  \right) e^{-\frac{1}{2}(\sigma t)^2} \right).
		\end{equation}
	\end{widetext}
	The model ~\eqref{eq:ZFPolarizationFit} consists of an anisotropic magnetic contribution
	characterized by an oscillating ``transverse'' component and a slowly relaxing ``longitudinal''
	component.  The longitudinal component arises due to the parallel orientation of the muon spin
	polarization and the local magnetic field. In polycrystalline samples with randomly oriented fields
	this results in a so-called ``one-third tail'' with $f_L = \frac{1}{3}$. For single crystals, $f_L$
	varies between zero and unity as the orientation between field and polarization changes from being
	parallel to perpendicular. Note that the whole volume of the sample is magnetically ordered nearly up to $T_{\rm C}$. Only very near the transition, there is a PM signal component, in addition to the magnetically ordered contribution, that is characterized by the densely distributed network of nuclear dipolar moments with a corresponding relaxation rate $\sigma$. The temperature-dependent fraction $0
	\le F \le 1$ governs the trade-off between magnetically-ordered and PM behaviors.\\

\textbf{Analysis of ZF-${\mu}$SR data under pressure}: In the pressure dependent experiments, a substantial fraction of the ${\mu}$SR asymmetry originates
from muons stopping in the MP35N pressure cell surrounding the sample.
Therefore, the ${\mu}$SR data in the entire temperature range were analyzed by
decomposing the signal into a contribution of the sample and a contribution of the pressure cell:
\begin{equation}
A(t)=A_S(0)P_S(t)+A_{PC}(0)P_{PC}(t),
\end{equation}
where $A_{S}$(0) and $A_{PC}$(0) are the initial asymmetries and $P_{S}$(t) and $P_{PC}$(t)
are the normalized muon-spin polarizations belonging to the sample and the pressure cell, respectively.
The pressure cell signal was analyzed by a damped Kubo-Toyabe function \cite{MaisuradzePC}.\\

\textbf{DFT calculations} Fully-relativistic band structure calculations were performed within the VASP \cite{VASP1,VASP2} code employing the projector-augmented wave method (PAW) \cite{PAW1,PAW2} and the Perdew, Burke, and Ernzerhof generalized-gradient (GGA-PBE) exchange-corellation functional \cite{PBE}. The AHC was calculated by means of the Wannier interpolation scheme \cite{AHC-wan}, as implemented in the Wannier90 package \cite{wannier90-1,wannier90-2}. The starting projections for constructing maximally-localized Wannier functions \cite{MLWF} were chosen as $sp3$ orbitals of Sn and S and the $d$-orbitals of Co. The upper bound for the inner and outer window for the band disentanglement procedure \cite{disent} were chosen at +2 and +5 eV relative to the Fermi level. The surface density of states was calculated for the Sn-terminated surface on the basis of the tight binding model derived from Wannier functions, by means of the WannierTools package \cite{WU2017}\\


\section{Acknowledgments}~
The ${\mu}$SR experiments were carried out at the Swiss Muon Source (S${\mu}$S) Paul Scherrer Insitute, Villigen, Switzerland using the high field HAL-9500 ${\mu}$SR spectrometer (${\pi}$E3 beamline), GPS instrument (${\pi}$M3 beamline) and high pressure GPD instrument (${\mu}$E1 beamline). 
The neutron diffraction experiments were performed at the Swiss spallation neutron source SINQ (HRPT and DMC diffractometers), Paul Scherrer Institute, Villigen, Switzerland. Z.G. thanks Vladimir Pomjakushin for invaluable
support with neutron diffraction experiments/analysis and for his useful discussions. H.C.L. thanks the support by the National Key RandD Program of China (Grants No. 2016YFA0300504), the National Natural Science Foundation of China (No. 11574394, 11774423, 11822412). This project has received funding from the European Research Council (ERC) under the European Union's Horizon 2020 research and innovation programm (ERC-StG-Neupert-757867-PARATOP). The magnetization measurements were carried out on the PPMS/MPMS devices of the Laboratory for Multiscale Materials Experiments, Paul Scherrer Institute, Villigen, Switzerland. This work was supported by the Swiss National Science Foundation (grant no. 206021{\_}139082).\\

%

\end{document}